\documentclass[a4paper,oneside]{aa}
\usepackage[T1]{fontenc}
\usepackage[latin1]{inputenc}
\usepackage{graphicx}
\usepackage{amssymb}
\usepackage[authoryear]{natbib}

\makeatletter

\providecommand{\tabularnewline}{\\}

\def\aj{AJ}                   
\def\araa{ARA\&A}             
\def\apj{ApJ}                 
\def\apjl{ApJ}                
\def\apjs{ApJS}               
\def\apss{Ap\&SS}             
\def\aap{A\&A}                
\def\aaps{A\&AS}              
\def\mnras{MNRAS}             
\def\pasp{PASP}               
\def\pasj{PASJ}               

\newcommand\myLabelTag{ R}
\newcommand\enumerateR{%
  \ifnum \@enumdepth >\thr@@\@toodeep\else
    \advance\@enumdepth\@ne
    \edef\@enumctr{enum\romannumeral\the\@enumdepth}%
      \expandafter
      \list
        \csname label\@enumctr\endcsname
        {\usecounter\@enumctr\def\makelabel##1{\hspace\labelsep \myLabelTag##1}}%
  \fi}

\makeatother
\begin{document}
\newcommand{\rstar}{early-R star}
\newcommand{\Rstar}{Early-R star}
\newcommand{\rstarabbrev}{early-R}
\newcommand{\Rstarabbrev}{Early-R}

\newcommand{\refchange}[1]{}

\newcommand{\posta}{n}

\title{Origin of the early-type R stars: \\
a binary-merger solution to a century-old problem\refchange{?}}
\author{R.G.~Izzard\thanks{email: R.G.Izzard@phys.uu.nl}\inst{1,3} \and C.S.~Jeffery\inst{2} \and J.~Lattanzio\inst{3}}
\institute{Astronomical~Institute~Utrecht, Postbus~80000, 3508~TA~Utrecht, The~Netherlands  \and 
Armagh Observatory, College Hill, Armagh BT61 9DG, Northern Ireland\and 
School of Mathematical Sciences, PO Box 28M, Monash University, Victoria 3800, Australia
}
\date{ }
\abstract{

The {\rstar}s are carbon-rich K-type giants. They are enhanced in
$^{12}\mathrm{C}$, $^{13}\mathrm{C}$ and $^{14}\mathrm{N}$, have
approximately solar oxygen, magnesium isotopes, $s$-process and iron
abundances, have the luminosity of core-helium burning stars, are
not rapid rotators, are members of the Galactic thick disk and, most
peculiarly of all, are all single stars. Conventional single-star
stellar evolutionary models cannot explain such stars, but mergers
in binary systems have been proposed to explain their origin.

We have synthesized binary star populations to calculate the number
of merged stars with helium cores which could be {\rstar}s. We find
many possible evolutionary channels. The most common of which is the
merger of a helium white dwarf with a hydrogen-burning red giant branch
star during a common envelope phase followed by a helium flash in
a rotating core which mixes carbon to the surface. All the channels
together give ten times more {\rstar}s than we require to match recent
Hipparcos observations -- we discuss which channels are likely to
be the true {\rstar}s and which are not. For the first time we have
constructed a viable model of the {\rstar}s with which we can test
some of our ideas regarding common envelope evolution in giants, stellar
mergers, rotation, the helium flash and the origin of the {\rstar}s.

}


\maketitle

\section{Introduction and History}

\label{sec:Introduction}The {\rstar}s are one of several classes
of carbon star, all distinguished spectroscopically by the presence
of strong absorption due to an excess of molecular carbon. While many
of these carbon stars are now relatively well understood, the {\rstar}s
have defied satisfactory explanation for nearly one hundred years.
In a nutshell, the problem is to explain how a single star which is
a giant, and hence has finished core-hydrogen burning, but is not
sufficiently luminous to have completed core-helium burning, can have
a surface which is contaminated by excess carbon. It may be argued
that such a straightforward statement of the problem is too simplistic;
it probably is. Therefore it is necessary to review here the main
characteristics of the {\rstar}s, to define the problem in more detail,
and thence introduce one approach to its solution.

\subsection{Spectral class definition}

\label{sub:Spectral-class-definition}Stars with peculiar banding
in their spectra were identified by \citet{1868MNRAS..28..196S} and
were later identified as carbon stars by \citet{1916POMic...2..103R}.
They are easily identified by strong absorption features due to $\mathrm{C}_{2}$,
CN or CH molecules. A subset of these, the R-type stars, was first
classified by \citet{1908HarCi.145....1F} based on the observations
of \citet{1896HarCi...9....1F}. The R stars {}``contain rays of
much shorter wave length than ordinary fourth type stars'' which,
in modern parlance, means they are bluer and hotter ($T_{\mathrm{eff}}\gtrsim4000\,\mathrm{K}$,
similar to the K-type stars, e.g. \citealp{1984ApJS...55...27D})
than the normal N-type carbon stars (with $T_{\mathrm{eff}}\lesssim3500\,\mathrm{K}$)
which are probably asymptotic giant branch (AGB) stars \citep{1983ARA&A..21..271I}.
The R stars are very common, apparently accounting for 1\% of all
K and G giants \citep{1984ApJS...55...27D}, and are 10 times more
numerous than N stars according to \citet{1965gast.conf..241B}.

Later, as part of the Henry Draper catalogue \citep{1918AnHar..91....1C}
and subject to revision by \citet{1928LicOB..13..123S}, the R class
was split into R0 to R8, where R0-4 (the hot/early/warm-R stars) are
warm, equivalent to normal K-type stars, while the cool/late-R stars
of type R5-8 are the carbon-rich equivalent of M stars. All R and
N stars were merged into a single type C by \citet{1941ApJ....94..501K},
but this was a {}``retrogressive step'' according to \citet{1972MNRAS.159..403E},
who reclassified late-R stars as N stars and left the {\rstar}s as
a distinct class. Other types of carbon stars complicate the issue,
especially in binaries where mass transfer can pollute the secondary
star with enough carbon to turn it into a dwarf carbon star or CH
star (see e.g. \citealp[sections 3 and 5]{1998ARA&A..36..369W}).
The J stars, which are similar to N stars with enhanced $^{13}\mathrm{C}$,
are also of uncertain origin but are clearly redder and more luminous
than the {\rstar}s \citep{2000ApJ...536..438A}. Example spectra
of all the carbon star types can be found in the atlas of \citet*{1996ApJS..105..419B}.
There is an alternative classification scheme developed by \citet*{2002A&A...390..967B}
who identify R0-3 stars as members of their HC1-HC3 class (where HC
stands for \emph{hot carbon}).

Standard single-star evolutionary models of AGB stars undergoing third
dredge-up (e.g. \citealp*{Parameterising_3DUP_Karakas_Lattanzio_Pols})
correctly, if sometimes only qualitatively, predict most of the properties
of the late-R and N type carbon stars. The {\rstar}s remain a distinct
group of stars with an unknown origin, just as they were in 1908.
More information about the carbon star family is given by \citet{2003PASA...20..314A},
although also useful are the reviews of \citet{1986MNRAS.220..723E},
\citet{1998M&PS...33..871L} and \citet{1998ARA&A..36..369W}.

\subsection{Spectroscopic Studies}

\label{sub:Spectroscopic-Studies}Spectroscopic studies of the {\rstar}s
have shed some light on their origin. The most comprehensive analysis
is that of \citet{1984ApJS...55...27D}. He found that {\rstar}s,
despite having $\mathrm{C}/\mathrm{O}>1$, have a low $^{12}\mathrm{C}/^{13}\mathrm{C}$
ratio (similar to the J stars), a solar or slightly sub-solar iron
abundance, enhanced nitrogen relative to solar, a solar oxygen abundance,
solar $^{17,18}\mathrm{O}/^{16}\mathrm{O}$ ratios and no $s$-process
enhancement. This is in contrast to N-type carbon stars, and late-R
stars, which show evidence of $s$-process elements, supporting the
idea that late-R stars are really just misclassified N stars. \citet{2006MmSAI..77..973Z}
confirm the findings of Dominy and support the idea that late-R and
N type stars are probably equivalent.\refchange{ Zamora (private
communication) also finds that {\rstar}s have $0.4\lesssim\log\epsilon(\mathrm{Li})\lesssim3.5$
-- perhaps indicative of some enhancement? -- while late-R stars have
$-0.5\lesssim\log\epsilon(\mathrm{Li})\lesssim0.5$, similar to G
and K giants \citep*{1980ApJ...235..114L}.} Late-R stars pulsate,
like N stars, but {\rstar}s do not, like K-type stars \citep*[fig. 8]{1996ApJS..105..419B,2002A&A...385...94B}.

The luminosity of {\rstar}s is around $100\mathrm{\, L_{\odot}}$,
typical of core helium burning (CHeB) stars (the red clump in a colour-magnitude
diagram, \citealp{1969MNRAS.144..449C,1970MNRAS.150..111C,1973ApJ...180..435F,1976ApJ...206..474S}),
while the luminosity of N and late-R stars is more than $1000\mathrm{\, L_{\odot}}$,
typical of AGB stars (\citealp{1958AJ.....63..477V}; \citealp{1974ApJ...190...85B};
\citealp{1976ApJ...206..474S}; \citealp{2002A&A...390..967B}).

\subsection{CH Stars}

\label{sub:CH-Stars}There remains the possibility that the {\rstar}s
are the CHeB counterparts of the CH stars in which, for some reason,
$s$-process elements are undetectable or not present. CH stars, both
dwarfs and giants, are formed by mass transfer to a main sequence
companion from an AGB star which has undergone third dredge-up. They
are rich in $s$-process elements and carbon which were formed in
the primary AGB star and are all binaries. The polluted secondary
evolves to the CHeB phase, perhaps remaining carbon-rich (although
see \citealp{2007astro.ph..2138S}). Metallicity affects the CH-formation
process in two ways. First, third dredge-up is more efficient at sub-solar
metallicities compared to solar \citep{Parameterising_3DUP_Karakas_Lattanzio_Pols}
so more primary carbon is made and transferred to the main sequence
star. Second, the initial abundance of oxygen is lower than at solar
metallicity so less carbon is required to be mixed into the stellar
envelope to form a carbon star (for which $C/O\geq1$ by number).
There is a threshold metallicity, around $Z\sim0.4Z_{\odot}$, above
which CH stars cannot form \citep{2002A&A...579...817}. Both these
facts are at odds with the properties of the {\rstar}s, which have
solar or slightly sub-solar iron and oxygen abundances \citep{1984ApJS...55...27D}.

\subsection{Binary Fraction}

\label{sub:Binary-Fraction}The nail in the AGB mass-transfer coffin
came with the work of \citet{1997PASP..109..256M} who found that
22 {\rstar}s -- \emph{all} those in his sample -- are \emph{single
stars}, even though $20\%$ of late-type giants are binaries. The
implication is that the {\rstar}s originate in binary mergers, as
it is hard to envisage a physical process which makes carbon stars
only in single stars and not in wide binaries. That they are single
and \emph{not} $s$-process enhanced has been used to distinguish
{\rstar}s from CH stars \citep{2003PASA...20..314A}. \citeauthor{1997PASP..109..256M}
also found that the {\rstar}s are not rapidly rotating, a feature
which must be explained by any potential {\rstar} model.

\subsection{Space Density and Distribution}

\label{sub:Space-Density-and-Distribution}\citet*{2001A&A...371..222K}
determined the absolute magnitudes and space density of the {\rstar}s
using Hipparcos data. They found that the {\rstar}s have magnitudes
similar to the red clump (i.e. CHeB) stars, the space density of the
{\rstar}s is $4.5-15\times10^{-8}\mathrm{pc}^{-3}$ and the R to
red clump number ratio is $0.04-0.14\%$. \citet{2002A&A...385...94B}
calculated the space density in the Galactic plane to be $1.66\times10^{-8}\,\mathrm{pc}^{-3}$
for {\rstar}s%
\footnote{Strictly, \citet{2002A&A...385...94B} define new spectroscopic types,
so we assume, as they state, that {\rstar}s are their HC' type, N
stars their CV type.%
}, $13.4\times10^{-8}\,\mathrm{pc}^{-3}$ for N stars, a ratio of $\sim16$.
The two papers give different number densities, but actually their
projected number densities are about the same. 

That the R stars are Galactic disk objects was recognised by \citet{1972MNRAS.159..403E}.
\citet{1984ApJS...55...27D} finds that N stars are younger disk objects
than, and distinct from, the R stars, while \citet{1960PASJ...12..214I},
\citet{1973PW&SO....1...1S} and \citet{1974A&A....33...21B} found
that the N and R stars are distributed differently across the sky,
with N stars more condensed in the Galactic plane. These results were
confirmed by \citet{2002A&A...385...94B} who found that R stars are,
on average, three times further from the Galactic plane. The solar
or slightly sub-solar iron abundance and velocity dispersion of the
R stars suggests they are members of the Galactic thick disk (\citealp{1944ApJ....99..145S,1958AJ.....63..477V,1964PASP...76..403D,1972MNRAS.159..403E};
\citealp{1994RMxAA..29..103K,2002A&A...385...94B}).

Third dredge up only occurs in solar-metallicity AGB stars above about
$1.3-1.5\mathrm{\, M_{\odot}}$ \citep*{1983MNRAS.202...59B,Parameterising_3DUP_Karakas_Lattanzio_Pols},
corresponding to a stellar lifetime of $5\,\mathrm{Gyr}$. Hence the
{\rstar}s cannot be intrinsic AGB stars if they are older than about
$5\,\mathrm{Gyr}$.

\subsection{Stellar Models}

\label{sub:Stellar-Models}The {\rstar}s pose a problem for stellar
evolution theory. Standard models of single stars are not carbon rich
except for thermally pulsing AGB stars and some massive Wolf-Rayet
stars. R-stars are too dim to be either of these, indeed their luminosities
are those of the CHeB (red clump) evolutionary phase. The problem
we face is twofold: when is carbon made in these stars and how do
we get it to the surface?

The binary star merger model satisfies most of the observational constraints.
We know there are no {\rstar}s dimmer than the red clump, so the
merger must lead to a carbon-rich, CHeB star. This naturally implies
that something odd happens immediately prior to the CHeB phase during
helium ignition at the tip of the giant branch. Canonical models of
the helium flash do not predict mixing of carbon-rich material from
the core to the stellar surface \citep{1966ApJ...145..496H}. However,
these models are one-dimensional and non-rotating, while a binary
merger leads to a three-dimensional, rotating system, so it is quite
possible that the merger and/or ignition leads to non-standard mixing.
One-dimensional models with parameterised rotational mixing and/or
low metallicity (with off-centre helium ignition) have led to some
mixing of carbon-rich core material with the stellar envelope (\citealp*{1976Ap&SS..41..407M,1977ApJ...216...57P,1990ApJ...351..245H}).
The models of \citet{1990ApJ...353..215I} simulate accretion of helium
onto a helium white dwarf (HeWD) and show that off-centre helium ignition
is a natural consequence of this process.

A number of two-dimensional simulations have been carried out by Deupree
and collaborators (\citealp{1980ApJ...239..284C,1981ApJ...247..607C,1987ApJ...317..724D,1996ApJ...471..377D})
some of which suggest the possibility of mixing material from a core
helium flash into the stellar envelope. It is hard to draw a conclusion
from these models as the results vary with both resolution and model
sophistication, and it is not clear how to apply the results to a
rapidly rotating binary merger. 

Recently, full three dimensional models of the helium flash have been
constructed (\citealp*{2006ApJ...639..405D}; \citealp{2006astro.ph.12147L})
although these are of single stars, not binary mergers (in the latter
paper \emph{slow} rotation was introduced). Simulations of HeWD mergers
have been constructed with a smooth particle hydrodynamics (SPH) approach
\citep*{2004A&A...413..257G}. While these do not evolve to the helium
flash, they do show that the cores lose very little mass, or angular
momentum, during their merger.

\subsection{Summary}

In this paper we will make, for the first time, a quantitative estimate
of the number of and properties of {\rstar}s using our binary population
nucleosynthesis model. We investigate possible channels for {\rstar}
formation by a binary merger process and subsequent helium flash in
a rapidly rotating core and also extrinsic (CH-star) channels by accretion
from a companion. We determine the effect of varying model parameters
on the {\rstarabbrev} to red clump and {\rstarabbrev} to N star
ratios, and compare these to observations. Section \ref{sec:Model}
describes our model, section \ref{sec:Results} our results, section
\ref{sec:Discussion} discusses the ramifications and we conclude
with some ideas for future research.

\section{Modelling the Mystery}

\label{sec:Model}We model populations of single and binary stellar
stars with the synthetic model of \citet*{2002MNRAS_329_897H}, updated
to include nucleosynthesis (\citealp{Izzard_et_al_2003b_AGBs}; \citealp{2006A&A...460..565I}
and changes outlined below). Our model follows the evolution of stars
using analytic fits to luminosity, radius, core mass etc. Interactions
with a companion due to tides, wind accretion and Roche-lobe overflow
(RLOF) are taken into account. Common envelope evolution is treated
with an $\alpha$-prescription \citep{1988ApJ...329..764L,1997MNRAS.291..732T},
where $\alpha$ is the fraction of the orbital energy which is transferred
to the envelope during the spiral-in phase. In most of our models
we use $\alpha=1$ and an envelope binding energy factor $\lambda=0.5$
\citep{1995MNRAS.273..146R}. Our synthetic AGB model is based on
the full evolutionary models of \citet*{Parameterising_3DUP_Karakas_Lattanzio_Pols}
and includes surface abundance changes due to third dredge-up and
hot-bottom burning. We also follow the surface abundances of massive
stars based on the models of \citet{2003MNRAS.341..299D}, as well
as yields due to supernovae and novae, but these are not important
with regard to the {\rstar}s.

\subsection{Stellar evolution}

\label{sub:Stellar-evolution}Our synthetic stars have solar-scaled
initial abundances according to \citet{1989GeCoA..53..197A}, with
metallicities $10^{-4}\leq Z\leq0.03$. There are many other parameters
in our model which are discussed in detail in \citet{2006A&A...460..565I}.
Where a parameter choice affects our results, we vary the parameter
within a reasonable range (see section \ref{sub:Variation-of-model-parameters}).
The main parameters which influence R-star formation are the metallicity
$Z$ and the common envelope parameters $\alpha$ and $\lambda$.
These are $Z=0.02$ (which we refer to as solar metallicity), $\alpha=1$
and $\lambda=0.5$ unless stated otherwise.

We have updated the \citet{2006A&A...460..565I} treatment of case-B
RLOF in binaries with primary masses between $0.75$ and $2.0\mathrm{\, M_{\odot}}$,
which turn out to be our prototype {\rstar}s. We follow the surface
abundances of $^{1}\mathrm{H}$, $^{4}\mathrm{He}$, $^{12}\mathrm{C}$,
$^{14}\mathrm{N}$ and $^{16}\mathrm{O}$ as a function of mass co-ordinate
based on detailed terminal main-sequence models constructed with the
TWIN stellar evolution code \citep{2002ApJ...575..461E}. As mass
is stripped during RLOF, transferred material is enhanced in helium
and nitrogen but deficient in carbon and oxygen because of CNO cycling
during the main sequence evolution of the primary. This is transferred
to the secondary star, enhancing its abundance of helium and nitrogen
while reducing hydrogen and carbon.

We modify the common envelope prescription of \citet{2002MNRAS_329_897H}
by removing their algorithm which instantaneously ignites merging
degenerate helium cores%
\footnote{The energy released in their algorithm usually disrupts the star,
which we consider unrealistic.%
}. Instead, in our model the less massive core is disrupted into a
disc around the more massive core and accretes until a single, rapidly
rotating core is formed \citep{2004A&A...413..257G}. The merged core
then grows by hydrogen-shell burning until helium ignites in a nuclear
runaway (the helium flash). For \emph{all our merged stars} we assume
that the ignition process, in a rapidly rotating core, mixes some
carbon into the stellar envelope as found by \citet{1977ApJ...216...57P}.
The star then settles into its CHeB phase as a{\posta} {\rstar}.

According to the \citet*{2000MNRAS.315..543H} formalism, helium ignites
when the stellar luminosity reaches a critical value which is a function
of the stellar mass and metallicity. This may not apply to our merged
stars which have abnormally large cores for their total mass and stage
of evolution, but given that we are ignoring the lifting of degeneracy
due to the merger we can probably do no better without detailed models
of {\rstar} progenitors. In our main merger channels (see below)
application of the \citet{2000MNRAS.315..543H} prescription leads
to minimum helium-ignition core masses of between $0.35$ and $0.45\mathrm{\, M_{\odot}}$,
compatible with the off-centre HeWD ignition models of \citet{1990ApJ...353..215I}
which have a core mass of $0.38\mathrm{\, M_{\odot}}$. In contrast,
an SPH simulation of the merger of two $0.4\mathrm{\, M_{\odot}}$
HeWDs does \emph{not} lead to a helium flash even though the maximum
temperature reached is $2\times10^{8}\,\mathrm{K}$ \citep*{2004A&A...413..257G}
-- any excess energy goes into lifting the degeneracy and expansion
of the white dwarfs. These simulations do not model the evolution
of the star beyond a few minutes after the collision, so it is possible
that the core contracts and/or increases in mass on a longer timescale,
but ignites while still rapidly rotating.

\subsection{Stellar population model}

\label{sub:Stellar-population-model}Our single star and binary primary
masses are distributed according to the initial mass function (IMF)
of \citet*{KTG1993MNRAS-262-545K} in the range%
\footnote{After some low-resolution trial runs the primary mass range was reduced
to $0.3\leq M/\mathrm{M_{\odot}}\leq8$ with little change to our
results.%
} $0.1\leq M/\mathrm{M_{\odot}}\leq80$, secondary star masses are
chosen from a distribution which is flat in $q=M_{1}/M_{2}$ such
that $0.1\mathrm{\, M_{\odot}}/M_{1}\leq q\leq1$ and initial separations
$a$ are chosen from a distribution flat in $\log a$ for $3\leq a/\mathrm{R}_{\odot}\leq10^{4}$.
We usually assume all binary orbits are circular ($e=0$) and \refchange{stars
evolve} to a maximum age of $13.7\,\mathrm{Gyr}$. Given these distributions,
the probability assigned to each binary system is $\Psi(M_{1},M_{2},a)$
and the contribution to the number of stars of a given type is $\int_{\mathrm{time}}\int_{M_{1},M_{2},a}S\times\delta(\mathrm{type})\times\Psi\, dt$,
where $\delta(\mathrm{type})=1$ when the star is of the required
type, $0$ otherwise. A similar calculation is performed for single
stars and the results combined with a $50\%$ binary fraction. We
set $S$, the star formation rate, to $1$ because we compare only
number ratios and relative number counts.

\subsection{Age selection criterion}

\label{sub:Age-selection-criterion}The {\rstar}s are associated
with the Galactic disk, particularly the thick disk \citep{2002A&A...385...94B}.
We select model stars older than the lower limit of the age-metallicity
relation for thick disk stars of \citet*{2004A&A...421..969B}, which
we fit to $5.61-6.68f\,\mathrm{Gyr}$ where $f=\max\left(-0.65,\left[\mathrm{Fe}/\mathrm{H}\right]\right)$.
At solar metallicity ($[\mathrm{Fe}/\mathrm{H}]=0$) this gives a
minimum age of $5.61\ \mathrm{Gyr}$, which is too old for intrinsic
carbon star formation according to our models \citep{Parameterising_3DUP_Karakas_Lattanzio_Pols}.

The imposition of the age limit is equivalent to ending star formation
$4.5\,\mathrm{Gyr}$ ago, which is not applicable to the whole Galactic
disk, even if it is correct for the thick disk. It implies, for example,
that there are \emph{no} intrinsic N type stars, which is incorrect
in the thin disk so we must be careful when comparing R to N star
number count ratios with the results of our model.

\subsection{N and red clump stars}

\label{sub:N-and-clump-stars}We define N-type carbon stars as giant
branch (GB) or AGB stars with $N_{\mathrm{C}}/N_{\mathrm{O}}\geq1$.
All our N stars are \emph{extrinsic:} there are \emph{no} intrinsic
carbon stars in our simulated populations because such stars are younger
than our $\sim4\mathrm{Gyr}$ age limit. 

The \emph{red clump} is synonymous with the CHeB phase of stellar
evolution (as defined by \citealp{2002MNRAS_329_897H}) in low-mass
stars, equivalent to the horizontal branch at higher mass.

\subsection{Early-R star formation channels}

\label{sub:R-star-formation-channels}The \citet{2002MNRAS_329_897H}
model defines many common-envelope merger channels which lead to a
CHeB star, any of which could be the {\rstar}s. We define our R-star
formation channels as follows: 

\begin{enumerateR}

\item Merger of a naked main-sequence helium star (HeMS) with a GB
star.

\item Merger of a core-helium burning star (CHeB) with a GB star.

\item Merger of a helium white dwarf (HeWD) with the core of a GB
star.

\item Accretion from an AGB companion, which is now a white dwarf
(at low metallicity, these are CH stars).

\item Merger of a HeWD with a Hertzsprung gap (HG) star which has
a partially degenerate helium core.

\item Merger of a GB star with a HG star.

\item Merger of two HG stars.

\item Merger of an AGB star with a helium-cored star. The helium
and CO cores mix completely to give a new, CO-rich helium core which
behaves as an evolved CHeB star. 

\item Merger of two naked helium stars.

\item Merger of a CO or ONe white dwarf (i.e. the core of an AGB
star) with an AGB star. This is similar to channel 8 but one star
has its envelope stripped.

\item Merger of two GB stars, similar to channel 3 but both stars
have a hydrogen envelope.

\end{enumerateR}

Our {\rstar}s are defined as those which have gone through one of
the above channels, are of the appropriate age (see section \ref{sub:Age-selection-criterion})
and are in the CHeB (red clump) phase. We assume that \emph{all} the
above mergers make {\rstar}s by the rotating core helium flash mechanism.

\section{Results}

\label{sec:Results}Table \ref{tab:population-fractions} gives the
relative formation rate and numbers of stars which form in each of
our model {\rstar} channels in our standard $Z=0.02$ population.
\begin{table}
\begin{centering}\begin{tabular}{|c|c|c|}
\hline 
Channel&
Formation Rate&
Number Fraction\tabularnewline
\hline
\hline 
R3&
$0.766$&
$0.705$\tabularnewline
R5&
$0.016$&
$0.041$\tabularnewline
R6&
$0.169$&
$0.196$\tabularnewline
R7&
$0.002$&
$0.005$\tabularnewline
R8&
$0.014$&
$0.004$\tabularnewline
R11&
$0.033$&
$0.048$\tabularnewline
\hline
\end{tabular}\par\end{centering}

\caption{\label{tab:population-fractions}The fraction of each merger channel
which contributes to our synthetic, standard $Z=0.02$ {\rstar} population
as defined in section \ref{sub:R-star-formation-channels}. The left
column \refchange{labels the channel} (R1\ldots{}11), the middle
column the relative formation rate for each channel and the right
column the relative number of {\rstar}s (assuming a constant star
formation rate and the \citealp{KTG1993MNRAS-262-545K} IMF). The
missing channels do not \refchange{contribute} to our R star population
and channel R4 is not a merger (we identify these with the CH stars
and treat them separately, see section \ref{sub:Channels-R3-and-R4}). }
\end{table}
The R3 (HeWD-GB) and R6 (HG-GB) channels together represent more than
$90\%$ of our {\rstar} progenitors -- we discuss these stars below
in some detail. We identify channel R4 with the CH stars. These form
in appreciable numbers at low metallicity, but not at all at solar
metallicity, as predicted by \citet{2002A&A...579...817} -- we discuss
these separately in section \ref{sub:The-Effect-of-metallicity}.
R5 (HeWD-HG) and R11 (GB-GB merger) possibly are {\rstar}s, forming
most of the remaining $10\%$. There are small contributions from
the R7 and R8 channels and none in the R1, R2, R9 and R10 channels.

\begin{figure*}
\begin{tabular}{cc}
\textbf{(a)}\includegraphics[bb=50bp 25bp 280bp 670bp]{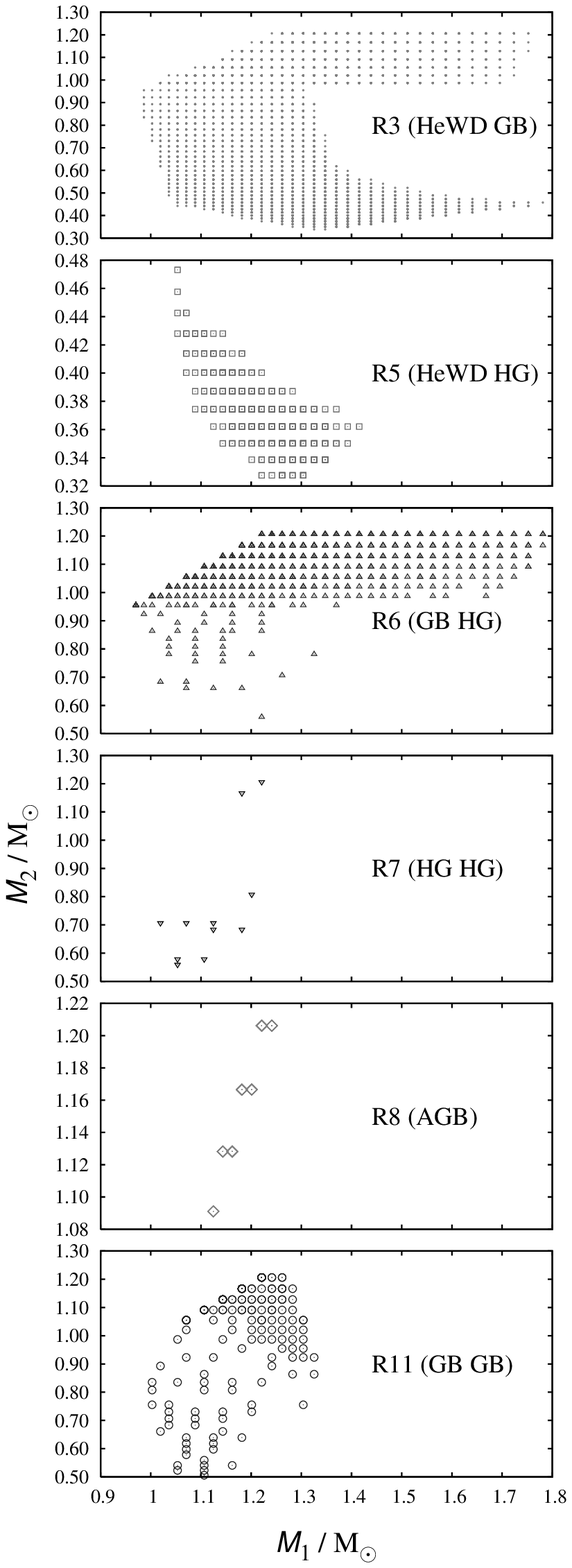}&
\textbf{(b)}\includegraphics[bb=50bp 25bp 280bp 670bp]{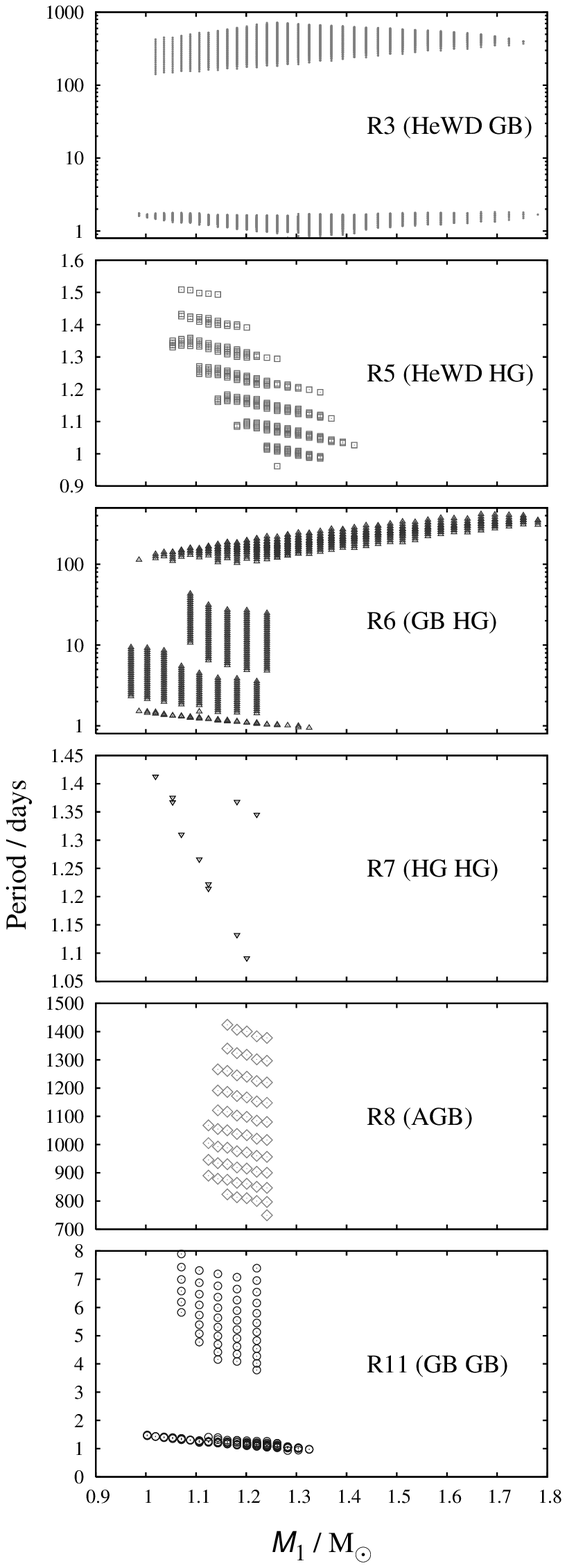}\tabularnewline
\end{tabular}

\caption{\textbf{\label{fig:Initial_distributions_R}(a)} Initial secondary
mass vs initial primary mass and \textbf{(b)} initial period vs initial
primary mass for all our {\rstar} merger progenitors (i.e. not including
channel R4, the CH stars). ($Z=0.02$, $\alpha=1$, $\lambda=0.5$.)
The symbols are as follows: R3 (HeWD-GB) small grey circles, R5 (HeWD-HG)
squares, R6 (GB-HG) upward-pointing triangles ($\bigtriangleup$),
R7 (HG-HG) downward-pointing triangles ($\bigtriangledown$), R8 (AGB)
large diamonds ($\diamond$), R11 (GB-GB) open circles. The contribution
from omitted channels is negligible. In panel (b) it is clear that
there are two distinct populations of R3, R6 and R11, the short period
binaries which have one common envelope phase, and the long period
binaries which have two.}
\end{figure*}

The parameter space of initial masses and periods which goes on to
form our {\rstar} candidates is shown in figure \ref{fig:Initial_distributions_R}.
\begin{figure*}
\hspace{-10mm}\begin{tabular}{cc}
\textbf{(a)}\includegraphics[scale=0.35,angle=270]{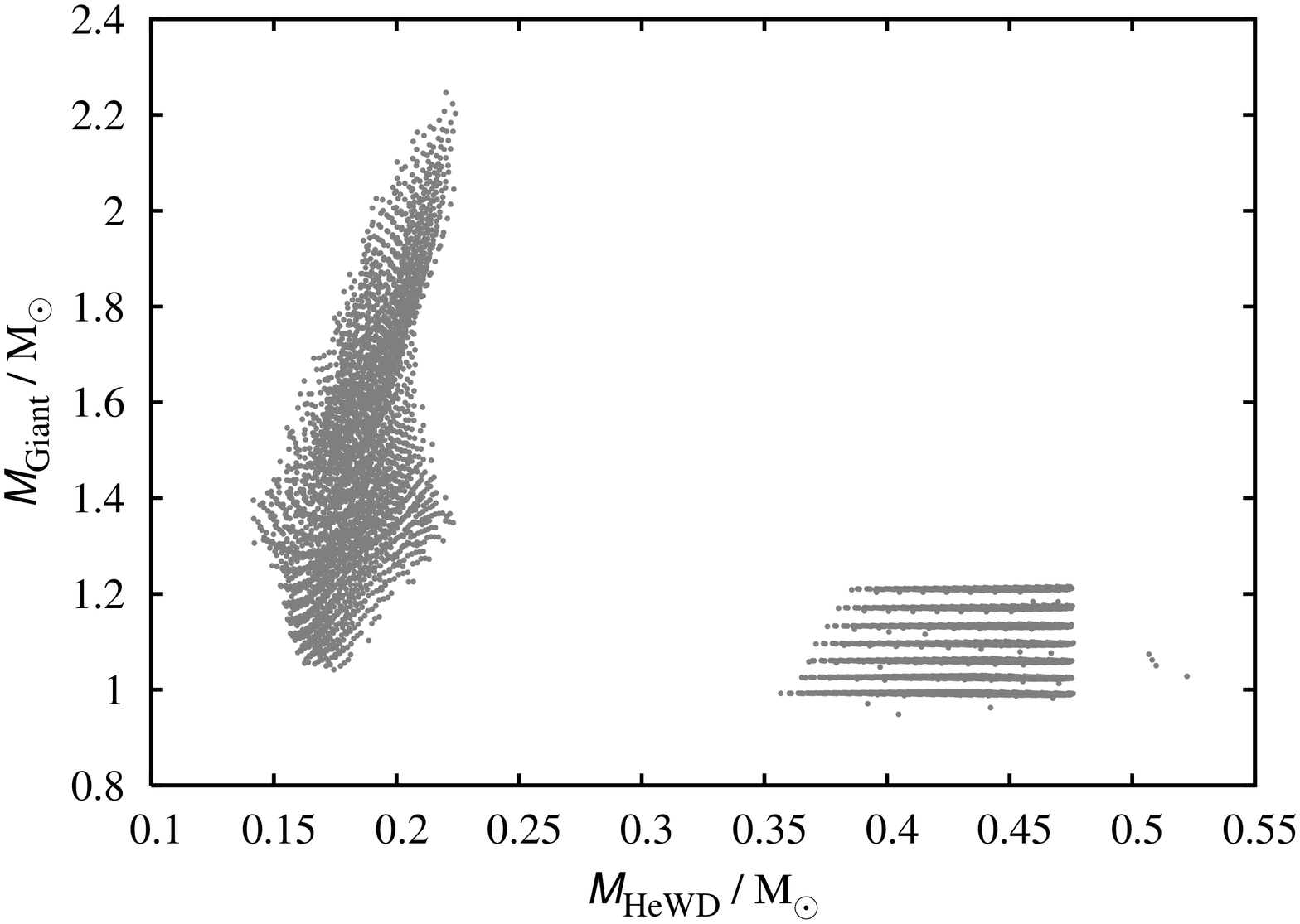}&
\textbf{(b)}\includegraphics[scale=0.35,angle=270]{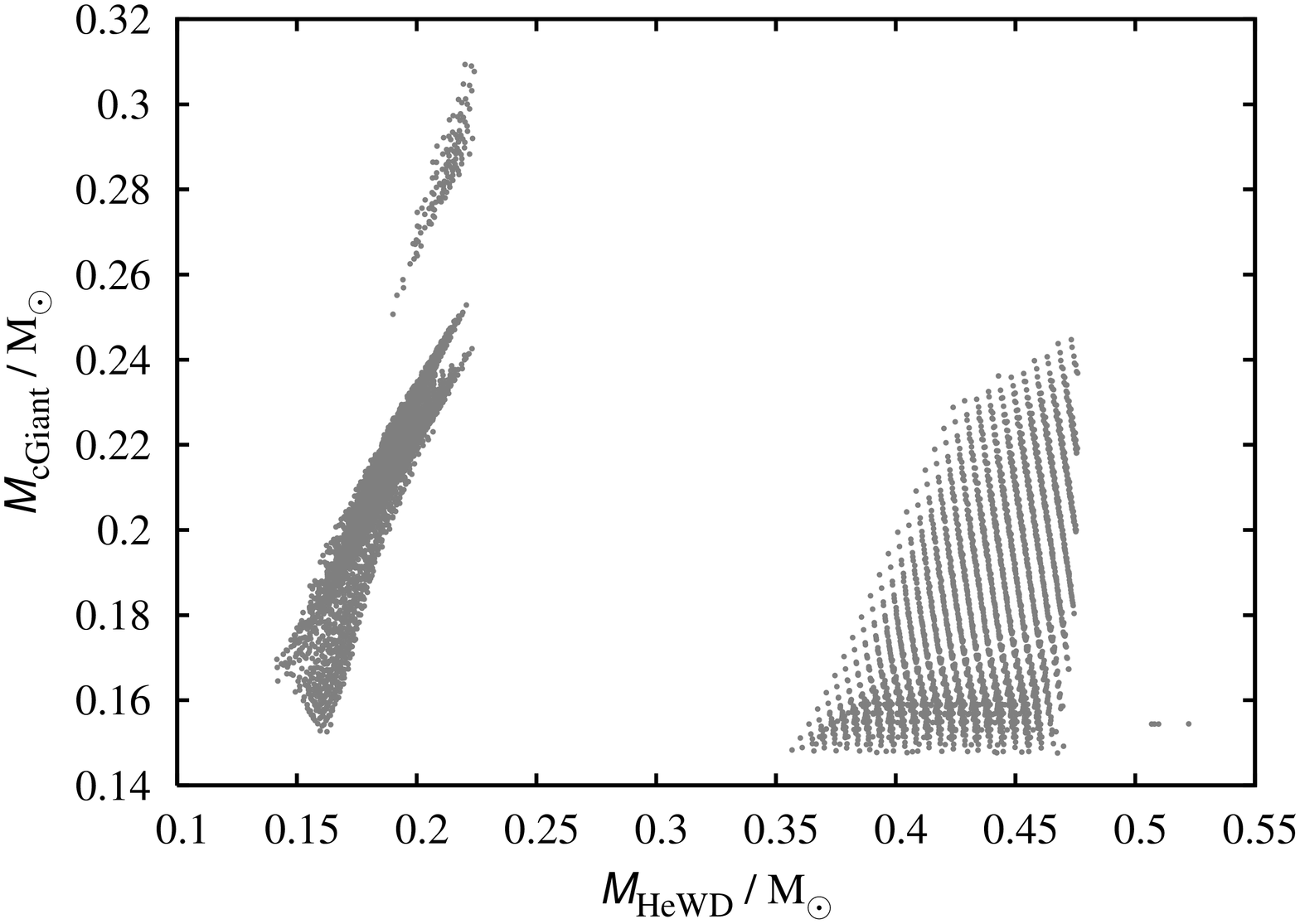}\tabularnewline
\end{tabular}

\caption{\label{fig:Pre_merger_distributions_R3}Pre-common envelope HeWD
mass vs giant mass \textbf{(a)} and giant core mass \textbf{(b)} for
our R3 channel. ($Z=0.02$, $\alpha=1$, $\lambda=0.5$.) In both
plots the group of points on the left side comes from our R3a (initially
short-period binary) models, while the points on the right are from
the (initially long-period) R3b channel. The R3b stars have such large
cores that they are probably too bright to be the observed R stars.}
\end{figure*}

\subsection{Subtype R3: HeWD + GB merger ($Z=0.02$)}

\label{sub:Subtype-R3:-HeWD-GB-merger}The R3 channel consists of
HeWD-GB mergers and dominates our {\rstar} progenitor population
at solar metallicity ($Z=0.02$). We have split the R3 type into subtypes
R3a and R3b representing initially short and long period binaries
as follows:

\begin{itemize}
\item \textbf{R3a} ($77\%$ of R3): $M_{1}=1-1.7\mathrm{\, M_{\odot}}$,
$M_{2}=0.3-1.0\mathrm{\, M_{\odot}}$, $P=1-2\,\mathrm{days}$ ($5\lesssim a/\mathrm{R_{\odot}}\lesssim8$)
~,
\item \textbf{R3b} ($23\%$ of R3): $M_{1}=1-1.7\mathrm{\, M_{\odot}}$,
$M_{2}=0.95-1.3\mathrm{\, M_{\odot}}$, $P=150-700\,\mathrm{days}$
($150\lesssim a/\mathrm{R_{\odot}}\lesssim450$)~.
\end{itemize}
Both the R3 subtypes originate from two phases of mass transfer, the
second of which is the common envelope phase which leads to merger
of two helium cores.

\subsubsection{R3a Sub-type}

\label{sub:R3a-Sub-type}In our R3a stars the first phase of mass
transfer is conservative RLOF when the more massive primary exhausts
its core hydrogen, crosses the Hertzsprung Gap (HG) and begins ascent
of the giant branch (GB). Initially, while $q<1$, the orbit shrinks,
but when enough mass is transferred that $q>1$ the orbit widens.
The primary completely loses its envelope and is left as a $\sim0.2\mathrm{\, M_{\odot}}$
HeWD \citep*[similar to the detailed models of][]{1967ZA.....66...58K}.
The secondary accretes a significant amount of mass and becomes a
blue straggler because it is more massive, hence bluer, than it should
be for its age. It is also nitrogen and $^{13}\mathrm{C}$ rich because
it accretes material which is stripped from CN-processed layers deep
inside the primary. The separation of these binaries, prior to the
evolution of the secondary up the giant branch, is about $20\mathrm{\, R_{\odot}}$.

The R3a channel contains the majority, $77\%$, of the R3 channel
stars because they have a higher formation rate and they have smaller
cores than R3b so spend longer in the CHeB phase.

\subsubsection{R3b Sub-type}

\label{sub:R3b-Sub-type}The R3b stars begin their evolution in a
relatively wide binary. When the primary evolves onto the GB it overflows
its Roche lobe and a common envelope results. During the spiral-in
of the helium core and main sequence star, the envelope of the giant
is lost, leaving a $0.38-0.48\mathrm{\, M_{\odot}}$ HeWD with a $1-1.2\mathrm{\, M_{\odot}}$
main sequence (MS) star in a reasonably close ($a\sim20\mathrm{\, R_{\odot}}$)
binary. The HeWD is about twice as massive in this scenario as compared
to R3a because the giant evolves further up the giant branch before
the first mass transfer. Note that in this scenario, the secondary
accretes only a small amount of mass (typically $\sim0.01\mathrm{\, M_{\odot}}$)
so while technically it is a blue straggler, it might not be detected
as such. Also, the secondary does not accrete much $^{13}\mathrm{C}$
or $^{14}\mathrm{N}$ from the stripped primary.

\subsubsection{Formation of R3 stars}

\label{sub:Formation-of-R3-channel-stars}In both the R3a and R3b
channels, after the first mass transfer the secondary evolves on its
nuclear timescale and eventually exhausts its core hydrogen. As it
crosses the HG and ascends the GB, RLOF begins and common envelope
evolution results.

Figure \ref{fig:Pre_merger_distributions_R3} shows the distribution
of masses and core masses just prior to the final common-envelope
phase and core merger. In both panels of the figure the R3a stars
are on the left and the R3b on the right. The giant core and the HeWD
merge during the ensuing common envelope phase to form a new single
\emph{}GB star with a core of $0.3-0.45\mathrm{\, M_{\odot}}$ for
the R3a and $0.5-0.7\mathrm{\, M_{\odot}}$ for the R3b channel. The
post-merger object is not immediately a{\posta} {\rstar}, but is
a GB star with an abnormally large core. In the case of R3b and more
massive R3a stars, the core is massive enough that the star very quickly
ignites helium. In the lowest-mass R3a stars some hydrogen shell burning
drives the core mass up to a minimum of $0.36\mathrm{\, M_{\odot}}$
when helium ignites. Once helium ignition has started, the luminosity
of the star drops and it settles into the red clump as a candidate
{\rstar}. 

Because the R3b stars have massive cores they have a mean luminosity
of about $650\mathrm{\, L_{\odot}}$. This is rather large, both compared
to the observed {\rstar}s and to our R3a channel, which has a mean
luminosity of about $170\mathrm{\, L_{\odot}}$. Once the core-helium
burning R-star phase is complete, these stars ascend the AGB, possibly
as J-type stars (see section \ref{sub:After-the-R-star}).

\subsection{The other merger channels ($Z=0.02$)}

\label{sub:The-other-merger-channels}The other merger channels arise
in much the same way as R3, with initially close binaries undergoing
conservative mass transfer onto a companion and initially wide binaries
passing through two common-envelope phases. These other phases are
considerably rarer because the initial binary parameters must be just
right such that the second mass-transfer phase occurs e.g. during
the HG rather than GB phase. Consequently, channels R5 (HeWD-HG),
R7 (HG-HG), R8 (AGB) and R11 (GB-GB) are rare compared to R3. The
R8 (AGB-merger) channel is almost certainly not related to the {\rstar}s
because the resulting merged stars have luminosities around $10^{3}\mathrm{\, L_{\odot}}$,
typical of stars with evolved CO cores (late-R stars or N stars).
We make no R9 or R10 stars at solar metallicity and although some
R10 (CO WD-GB or ONe WD-GB mergers) systems do exist at lower metallicity
their numbers are very few. We never make any R1 or R2 systems because
we select only old stars and potential R2 stars merge as GB-GB systems
(i.e. prior to the GB-CHeB phase).

Channel R6, the merger of a GB star with a HG star, is relatively
common. Most of these systems ($85\%$) are similar to the R3a systems
but with an initially more massive secondary, such that it is a HG
star (rather than a MS star) when the primary overflows its Roche
lobe. As a result, a common envelope forms during the \emph{first}
mass transfer phase and the cores merge.

The remainder of the R6 systems, and most of the R11 systems, are
initially wide binaries containing a HeWD and a HG (or GB, in the
case of R11) star. The wind from the HG star is accreted onto the
HeWD at a rate sufficient to form a new envelope and rejuvenate the
HeWD as a GB star. A common envelope phase soon follows and the HeWD
merges with the HG star. These systems suffer the same problem as
the R3b channel, their cores are massive ($0.6-0.8\mathrm{\, M_{\odot}}$)
and they are are too luminous ($\gtrsim600\mathrm{\, L_{\odot}}$)
to be {\rstar}s.

\subsection{Channel R3 vs R4 (CH) as a Function of Metallicity}

\label{sub:Channels-R3-and-R4}%
\begin{figure}
\begin{centering}\includegraphics[bb=50bp 98bp 554bp 770bp,scale=0.36,angle=270]{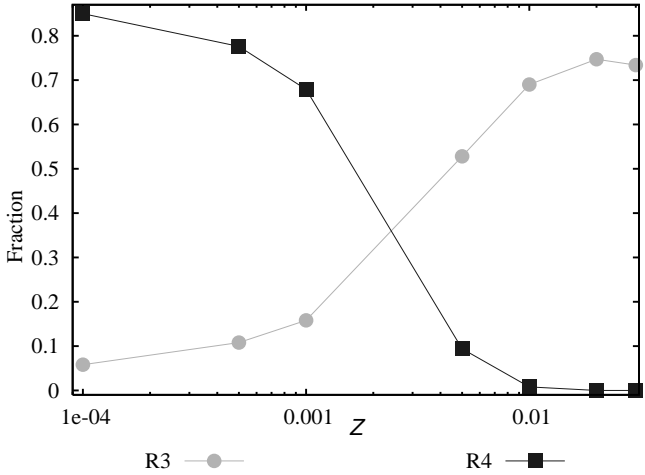}\par\end{centering}

\caption{\label{fig:r3r4fractions}Our model predictions for the relative
fractions of the R3 and R4 channels as a function of metallicity.
While R4 dominates at low metallicity (these are halo CH stars), channel
R3 (HeWD-GB mergers) dominates at around solar metallicity~--~these
are our prime candidates for the {\rstar}s.}
\end{figure}
Figure \ref{fig:r3r4fractions} shows the relative number of the R3
and R4 channels as a function of metallicity. The CH-star channel
R4 dominates our {\rstar} population for $Z\lesssim0.004$ while
above this metallicity the binary merger channel R3 is predominant.
The reason for this is a combination of increasing oxygen abundance
(proportional to $Z$) and decreasing third dredge-up efficiency,
as discussed in section \ref{sub:CH-Stars}. Our conclusion is that
for $Z\lesssim0.004$, e.g. in the Galactic halo, most carbon-rich
red clump stars are probably CH stars, with enhanced $s$-process
abundances and a binary companion, while at higher metallicity (in
the Galactic disk) the merger model dominates and all carbon-rich
red clump stars should be single {\rstar}s. This is just as is observed
(see section \ref{sub:R/CH-stars-metallicity-population}).

\subsection{The Effect of Metallicity on Merger Channels}

\label{sub:The-Effect-of-metallicity}%
\begin{figure}
\begin{centering}\includegraphics[scale=0.36,angle=270]{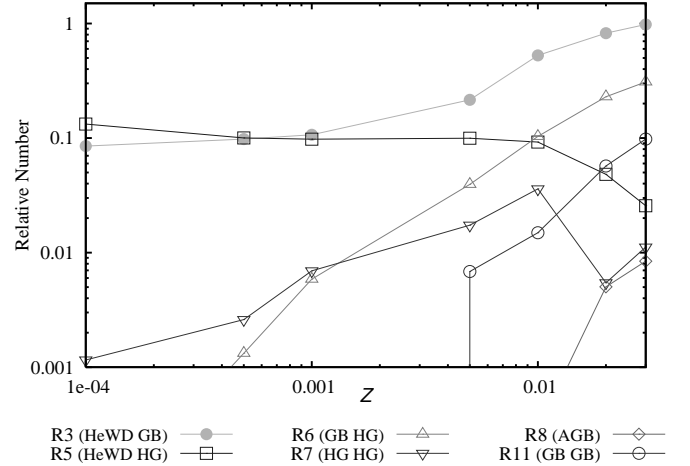}\par\end{centering}

\caption{\label{fig:merger-fractions}The relative number of our {\rstar}
merger candidates as a function of metallicity, arbitrarily normalized
to $1$ for the R3 channel at $Z=0.03$. The frequency of most merger
types decreases as the metallicity drops. The symbols are the same
as in figure \ref{fig:Initial_distributions_R}.}
\end{figure}
Figure \ref{fig:merger-fractions} shows the effect of metallicity
on the relative number of each merger channel. Lowering the metallicity
reduces the number of mergers of all types mainly because of our age-selection
criterion (see section \ref{sub:Age-selection-criterion}). The R5
channel (HeWD-HG) increases because as the metallicity drops the stellar
lifetime for a given mass decreases, so lower mass stars evolve off
the main sequence faster than they would at solar metallicity. The
IMF favours low-mass stars, hence the increase in R5.

\subsection{Early-R to red clump ratio}

\label{sub:R-to-red-clump-ratio}The ratio of the number of R to red
clump stars was determined to be $0.04-0.14\%$ by \citet{2001A&A...371..222K}.
\begin{figure}
\begin{centering}\includegraphics[bb=50bp 80bp 554bp 770bp,scale=0.36,angle=270]{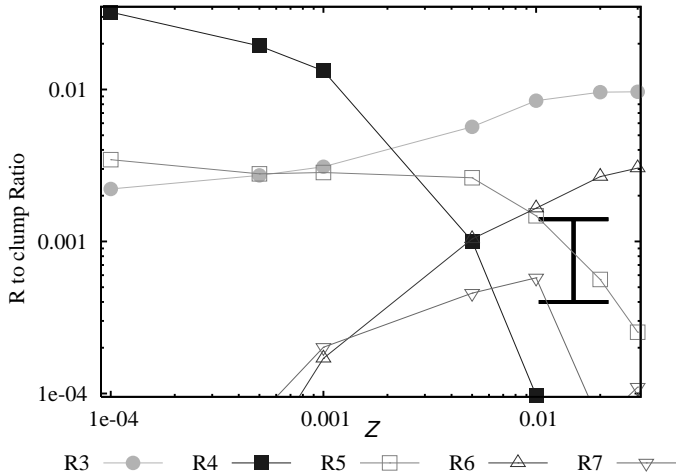}\par\end{centering}

\caption{\label{fig:r_to_clump_ratio}Our calculated R to CHeB (red clump)
number ratio as a function of metallicity for channels R3 to R7. The
other channels are negligible. The symbols are the same as in figure
\ref{fig:merger-fractions}. The thick error bar shows the observational
bounds from \citet{2001A&A...371..222K} for the Galactic disk. The
R3, 5, 6 and 7 subtypes could all be the real {\rstar}s, although
in the case of R3 and R6 only a subset of our model stars can be true
{\rstar}s because we make too many.}
\end{figure}
 Figure \ref{fig:r_to_clump_ratio} compares the results of our models
to this ratio and it is obvious that our model overestimates the number
of {\rstar}s (or underestimates the number of CHeB stars). In our
R-star selection criteria we have tried to be as inclusive as possible,
so (for example) if only $10\%$ of our R3 stars go on to mix carbon
into their envelopes when they ignite helium, then our models would
agree with the observations -- we discuss this further in section
\ref{sec:Discussion}. It is unlikely that we have underestimated
the number of CHeB stars by a factor of ten.

\subsection{The Early-R to N ratio }

\label{sub:R-to-N-ratio}The ratio of the number of R to the number
of N stars is about 10 \citep{1984ApJS...55...27D}. Our models give
$N_{\mathrm{R}}/N_{\mathrm{N}}\sim10-100$, but this is spurious because
we do not include young, intrinsic N stars and some of the R stars
used to calculate the observed ratio are probably \emph{cool/late}-R
(i.e. N type). Also, the observed ratio depends very strongly on Galactic
latitude (e.g. \citealp{1944ApJ....99..145S}, compare Figs 1 and
2) and/or height above the Galactic plane \citep{2002A&A...385...94B}.
To test this one would build a Galactic population model, with separate
thick and thin disk populations, but this is beyond the scope of this
paper.

On the other hand, our model results are not inconsistent with the
observations. We can estimate the ratio $N_{\mathrm{R}}/N_{\mathrm{CHeB}}$
from the observed $N_{\mathrm{R}}/N_{\mathrm{N}}$ by calculating
$N_{\mathrm{R}}/N_{\mathrm{N}}\times N_{\mathrm{N}}/N_{\mathrm{CHeB}}$.
\citet{2002A&A...385...94B} give $N_{\mathrm{R}}/N_{\mathrm{N}}\sim0.32$
in the Galactic plane%
\footnote{Again, we associate their type CV stars with N stars, and their HC'
stars with {\rstar}s.%
}, while we can approximate $N_{\mathrm{N}}/N_{\mathrm{CHeB}}$ from
$\Delta t_{\mathrm{N}}/\Delta t_{\mathrm{CHeB}}\sim2/150=0.013$ where
the $\Delta t$s are the lifetimes (in $\mathrm{Myr}$) of the AGB
and CHeB evolutionary phases for a $Z=0.02$, $1.3\mathrm{\, M_{\odot}}$
star. This results in $N_{\mathrm{R}}/N_{\mathrm{CHeB}}\sim0.4\%$,
similar to the $0.1\%$ \citet*{2001A&A...371..222K} find and within
the range of our results.

\subsection{Early-R/GK giant ratio}

\label{sub:R/GK-giant-ratio}The R to GK giant ratio is 1\% according
to \citet{1984ApJS...55...27D}, although the source of this number
is not given (it is presumably calculated from the carbon star catalogue
of \citealp{1973PW&SO....1...1S} and probably includes late-R stars).
If we define GK giants as GB and AGB stars with $3800\leq T_{\mathrm{eff}}/\mathrm{K}\leq5850$
which satisfy our age criterion, which is quite conservative as we
should probably include CHeB stars as well, then at solar metallicity
($Z=0.02$) our {\rstarabbrev} to GK giant ratio is about $0.5\%$.
Given that our {\rstar} count is as high as we can possibly make
it (it should be a factor of ten less to match the R to red clump
ratio) and we do not include younger GK giants, our R to GK giant
ratio is clearly different to that quoted by \citeauthor{1984ApJS...55...27D}.

\subsection{Dredge-up during the core helium flash}

\label{sub:Dredge-up-during-the-flash}
\begin{figure}
\begin{centering}\includegraphics[scale=0.35,angle=270]{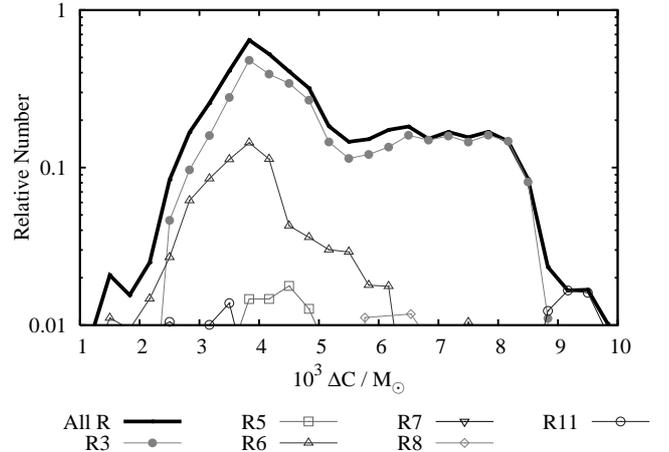}\par\end{centering}

\caption{\label{fig:DUP-required} Distribution of the mass of carbon which
must be dredged up in order to convert our merger stars into a carbon
star ($Z=0.02$ standard model). The thick black line is the sum of
all our R-merger channels, the lines with points are the individual
channels.}
\end{figure}

We postulate that, during the
core helium flash of our merged objects, there must be some dredge
up of carbon from the core into the envelope. We cannot model this
in detail, but we can estimate the minimum amount of carbon, $\Delta_{\mathrm{C}}$,
required to make~$\mathrm{C}/\mathrm{O}=1$. We find

\begin{eqnarray}
\Delta_{\mathrm{C}} & \approx & M_{\mathrm{env}}\frac{\left(\frac{3}{4}X_{\mathrm{O}}-X_{\mathrm{C}}\right)}{1-X_{\mathrm{C}}}\,,\end{eqnarray}
where $M_{\mathrm{env}}$ is the common envelope mass and $X_{\mathrm{C}}$
and $X_{\mathrm{O}}$ are the abundances of carbon and oxygen initially
present in the stellar envelope (the derivation of this formula is
in Appendix \ref{sub:Carbon-Dredge-Up}). In our $Z=0.02$ simulations
$\Delta_{\mathrm{C}12}$ is usually in the range $0.003$ to $0.01\mathrm{\, M_{\odot}}$,
as shown in figure \ref{fig:DUP-required}. %

\citet{1986ApJ...303..649D} finds a much smaller amount of carbon,
around $10^{-6}\mathrm{\, M_{\odot}}$, is mixed to the stellar envelope
in two-dimensional calculations of the helium flash. These simulations
were necessarily of low resolution, and did not include rotation,
which should be important in mixing material out of the core and through
the hydrogen-burning shell (\citealp{1976Ap&SS..41..407M} see also
our section \ref{sub:Merger-mixing}).

\subsection{Variation of model parameters}

\label{sub:Variation-of-model-parameters}The initial metallicity
is the parameter which most affects the formation of {\rstar}s by
the binary merger channel. Here we consider some other parameters
of our model which may affect the formation of {\rstar}s. 

\begin{itemize}
\item The common envelope parameter $\alpha$, determines the fraction of
the orbital energy which is used to eject the envelope. This most
affects the initially long-period binary channels which go through
two phases of common envelope evolution e.g. the R3b channel. If $\alpha$
is small the binary merges during the \emph{first} common envelope
phase, meaning the second never happens. This reduces the $Z=0.02$
R3 population by about $25\%$ (i.e. by the fraction which are R3b
stars) when changing from $\alpha=1$ to $\alpha=0.1$.
\item We have two prescriptions for $\lambda$, the giant envelope binding
energy factor. The first is to use a constant value of $0.5$, as
in \citet{2002MNRAS_329_897H} -- this is our standard model. Our
second prescription is based on the models of \citet{2000A&A...360.1043D,2001ASPC..229..255D}
who calculated $\lambda$ from detailed stellar evolution models.
With the variable-$\lambda$ prescription the R-star frequency is
reduced by a only few per cent relative to $\lambda=0.5$. 
\item The amount of third dredge up in AGB stars less massive than $1.5\mathrm{\, M_{\odot}}$
is a debated subject. Detailed stellar evolution models suggest no
dredge up in these stars, but extra dredge up is required in AGB stars
in the Magellanic clouds in order to match the observed carbon star
luminosity function \citep{1981ApJ...246..278I}. \citet{2006A&A...445.1069G}
claim we should not increase the dredge up, but Bona\v{c}i\'c Marinovi\'c
et al. (2007, A\&A submitted) confirm that extra dredge up is required
on the basis of $s$-process abundances. Regardless, we tested the
effect of increasing the amount of the third dredge up as a function
of metallicity according to the prescription of \citet{Izzard_et_al_2003b_AGBs}.
The extra dredge up increases the number of extrinsic carbon stars
and slightly increases the threshold metallicity at which the number
of R3 stars is equal to the number of R4 stars. At solar metallicity,
there are still no intrinsic carbon stars in our population and this
parameter has no effect.
\item Introducing the companion reinforced attrition process (CRAP) of \citet{1988MNRAS.231..823T}
has little effect on the number of {\rstar}s, reducing the number
of R3 stars by just a few per cent for large values of the CRAP parameter
($10^{4}$, as suggested in \citeauthor{1988MNRAS.231..823T}). However,
for such large values the R5 channel becomes important (about half
as many as R3) because enhanced mass loss on the giant branch increases
the number of HeWDs available for mergers. 
\item Changing the initial eccentricity, $e$, of the binary population
from $0$ to $0.75$, which simulates the effect of introducing a
distribution in $e$, has little effect on our results.
\end{itemize}
We have not considered changing the initial distributions such as
the IMF, separation distribution etc. Such an effort is beyond the
scope of this paper. Strictly, we should apply a Galactic metallicity
and star-formation rate distribution (perhaps as a function of height
above the Galactic plane) rather than compare single-population, constant
star-formation rate populations. Again, this is beyond the scope of
this paper but we hope to examine this problem in the future.

\section{Discussion and Hope}

\label{sec:Discussion}\refchange{Our common-envelope merger followed
by rotating core helium flash model makes a sufficient number of {\rstar}s
to match observations.} This raises hope of understanding these stars
at last, but many questions remain. Here we discuss the merits of
the different channels and consider the mixing and angular momentum
transport which may occur during the core merger.

\subsection{The R-channels ($Z=0.02$)}

\label{sub:The-R-channels}All our R channels together cannot be the
equivalent of the observed {\rstar}s because we make too many by
a factor of about ten. There are advantages and disadvantages to each
channel which, in our opinion, make some more likely candidates than
others.

\begin{description}
\item [{R3}] The HeWD-GB mergers. The major disadvantage of the initially
short-period R3a channel is the delay time, about $2.5\,\mathrm{Myr}$,
between the core merger and helium ignition. The equivalent in the
R3b channel is less than $1\,\mathrm{Myr}$. A lengthy delay may allow
transfer of material and/or angular momentum from the (pre-flash)
core to the envelope, which would slow the core rotation and reduce
the likelihood of mixing of carbon into the envelope during the helium
flash. \\
The advantage of the R3a channel is that the cores are of low enough
mass, mostly $\lesssim0.45\mathrm{\, M_{\odot}}$, to be degenerate
when helium ignites, provided they can lose any energy they gained
from the merger process (which would otherwise lift the degeneracy,
as in \citealp{2004A&A...413..257G}) and still remain rapidly rotating.
If $M_{\mathrm{c}}$ (the merged-core mass) is the parameter which
determines whether the helium flash mixes material into the envelope
then only a small range ($\sim0.02\mathrm{\, M_{\odot}}$ e.g. $0.38-0.40\mathrm{\, M_{\odot}}$)
in $M_{\mathrm{c}}$ selects the required $10\%$ of R3 stars. The
relatively high core mass of the R3b channel means its stars are many
times brighter than the observed R stars, with $\log L/\mathrm{L_{\odot}}\sim2.5-3.0$.
They also span a wider effective temperature range, from $\log T_{\mathrm{eff}}/\mathrm{K}=3.55$
to $3.9$, than the R3a stars which cluster at $\log T_{\mathrm{eff}}/\mathrm{K}\sim3.68$.
Possibly these would be observed as J stars or as A and/or F giants.\\
R3a stars, and our other R channels which involve significant accretion
from a giant companion, are enhanced in $^{13}\mathrm{C}$ and $^{14}\mathrm{N}$,
as seen in the {\rstar}s \citep{1984ApJS...55...27D}. These enhancements
may also be due to mixing across a hydrogen shell (see section \ref{sub:Merger-mixing}).
\item [{R4}] The binary R4 channel, equivalent to CH stars, cannot work
at solar metallicity (see section \ref{sub:R/CH-stars-metallicity-population}
below) and does not lead to single stars, so this is definitely \emph{not}
the channel which makes the {\rstar}s.
\item [{R5}] Channel R5, the HeWD-HG mergers, is similar to R3 but contributes
a factor of 17 fewer stars in our standard $Z=0.02$ model. Its main
advantages are that the number of {\rstar}s formed is similar to
that which is observed and that HG stars are less compact than GB
stars so more likely to keep their envelopes and merge in a common
envelope phase \citep{2006astro.ph.11043T}. The disadvantage is that
the core of a HG star may not be sufficiently degenerate for helium
to ignite soon after the merger so the core has time to lose its angular
momentum. On the other hand, the merged core mass is always low (about
$0.3\mathrm{\, M_{\odot}}$) because the HG star is relatively unevolved
compared to an equivalent mass GB star. Assuming the merged core can
maintain its rotation rate as the R5 star ascends the giant branch
these may be the {\rstar}s.
\item [{R6}] The GB-HG mergers occur at a rate around $1/4$ that of the
R3 channel -- still too many compared to the observations, but a better
match than R3. As in R5, the advantage of a HG star is that it is
likely to survive the common envelope phase, but they may not be degenerate
or massive enough to be still rotating when helium ignites. A fraction
of R6 forms from double common-envelope systems and, as in channel
R3b, these are more luminous than observed R stars.
\item [{R7}] These are the HG-HG mergers, which are very rare (150 times
fewer than the R3 channel) so are probably not the {\rstar}s.
\item [{R8}] These AGB-helium core mergers are also rare, with numbers
similar to R7. It is also not likely that the assumption of our model
-- that the helium core \emph{completely} mixes with the CO core of
the AGB star -- is true. In any case, these are too bright to be the
{\rstar}s (but may be J/N/late-R stars).
\item [{R11}] GB-GB mergers are rare, a factor of 15 less than the R3 channel,
but cannot be ruled out as potential {\rstar} progenitors. However,
they are the binaries most likely to lose their envelopes when they
merge. Both cores are degenerate and usually of low mass and often
one member of the binary is a rejuvenated WD with a very thin envelope,
so these are very similar to R3 mergers. Stars in this channel span
a wide range in luminosities (from $60\mathrm{\, L_{\odot}}$ to $900\mathrm{\, L_{\odot}}$)
and effective temperatures ($\log T_{\mathrm{eff}}/\mathrm{K}=3.53-3.9$).
In a few cases the post-merger object has a very thin envelope, so
is very blue ($T_{\mathrm{eff}}\sim40,000\mathrm{K}$).
\item [{R1,~2,~9~\&~10}] None or very few of these are made in our
populations.
\end{description}
Most likely only some fraction of our model {\rstar}s will ignite
and mix carbon into the envelope, but we do not know which stars these
are. Our population synthesis model predicts too many {\rstar}s.
This is a positive result, because we have tried to select every possible
merger channel. It is also an interesting challenge because we do
not know the true R-star formation channel.

\subsection{R/CH stars, metallicity and population}

\label{sub:R/CH-stars-metallicity-population}Above a threshold metallicity
$[\mathrm{Fe}/\mathrm{H}]\sim-0.3$ and in an exclusively old stellar
population, our models suggest that the {\rstar}s should be the \emph{only}
carbon stars. The Galactic bulge, which is old and of (super-)solar
metallicity \citep{2003A&A...399..931Z}, is an obvious target for
R star surveys. In fact, large numbers of low-luminosity carbon stars
have already been found in the bulge and have been identified as {\rstar}s
in the surveys of \citet*{1985A&A...145L...4A}; \refchange{\citet{1991A&AS...88..265A}
and \citet{1991A&A...244..367W}}. This is in contrast to the very
few or even total lack of  N-type stars in the bulge \citep{1998IAUS..184...11R},
in agreement with our results.

At metallicities typical of the Galactic halo, the number of CH stars
vastly exceeds the number of {\rstar}s. That is not to say there
are no {\rstar}s in the halo, because according to our models there
should be. With regard to the detection of {\rstar}s in the halo,
there are several reasons it may be difficult: 

\begin{enumerate}
\item There should be many more CH stars than {\rstar}s in the halo. At
$[\mathrm{Fe}/\mathrm{H}]=-2.3$ we predict $\mathrm{R}3/\mathrm{R}4\sim7\%$
which means the {\rstarabbrev}/CH ratio will be much smaller (R4
only includes core helium burning CH stars, the true CH population
will contain both these and giants and possibly dwarfs) especially
if we reduce our figure by a factor of ten to agree with the Hipparcos
observations.
\item Halo stars are rare compared to disk stars, so there are simply not
as many population-II {\rstar}s.
\item CH stars are likely to be selected by colour, which may omit {\rstar}s.
\item Giant branch CH stars are somewhat brighter than CHeB stars, so any
survey aimed at CH stars (rather than C dwarfs) must be quite sensitive
to relatively dim stars in order to detect a{\posta} {\rstar}.
\item Is it possible to distinguish easily between {\rstarabbrev} and (core-helium
burning) CH stars? According to \citet{2003PASA...20..314A} the only
differences are the presence of $s$-process isotopes and binarity,
data which are often difficult to obtain. As such, {\rstar}s \emph{already
found in the halo} could have been misclassified as CH stars. A recent
survey \emph{has} detected a few halo {\rstar}s and gives a method
for differentiating between {\rstarabbrev} and CH stars \citep{2005MNRAS.359..531G}.
\end{enumerate}
The \citet{1995A&A...303..107W} survey of the Large Magellanic Cloud
(see also \citealp{1993A&AS...97..603R}), which has a sub-solar metallicity
$[\mathrm{Fe}/\mathrm{H}]\sim-0.3$ \citep{2006A&A...448...77C},
finds a menagerie of N, R and J type stars, over a wide range of luminosities,
in qualitative agreement with our model population. We should caution,
however, that the LMC probably contains a significantly younger population
of intrinsic N type stars \citep{2006A&A...452..195C} which makes
a statistical comparison difficult because our population model does
not include young stars.

We also have not considered the impact of the latest determinations
of the solar oxygen abundance (\citealp*{2001Allende_etal,2005ASPC..336...25A})
which decreases the amount of oxygen by a factor of two relative to
\citet{1989GeCoA..53..197A}. This reduces our estimate of the mass
of carbon which must be mixed into the envelope during helium ignition,
$\Delta_{\mathrm{C12}}$, and shifts the metallicity at which CH stars
form, but does not significantly change our conclusions. 

Finally, we are not sure why Galactic {\rstar}s are located only
in the thick disk \citep{2002A&A...385...94B}. Some of their HC1-3
(R-type) stars are of low radial velocity, so could be members of
the thin disk. Alternatively, the initial binary fraction, and hence
number of mergers, may be higher in the thick disk compared to the
thin disk \citep{2006astro.ph.12172G}.

\subsection{Progenitors observed}

\label{sub:Progenitors-observed}The progenitors of our {\rstar}s
are binaries in which one star is a WD, the other a blue straggler
-- one such star has been observed \citep{1997ApJ...481L..93L}. Recently,
models of MS+WD blue straggler systems were constructed by \citet{2006A&A...455..247T}
but they cannot take their calculations through the final common envelope
phase. Regarding the MS+WD phase, however, they find quite similar
results to ours despite their use of full stellar evolution rather
than synthetic code.

Blue stragglers (BSs) are most easily identified in globular clusters
(GCs) because the stars in a particular cluster are of the same age
so BSs are easily seen to be brighter and bluer than the tip of the
main sequence. However, binary mergers may be rare because of the
low binary fraction in the core of clusters \citep{2005MNRAS.358..572I}
-- or perhaps the binary fraction is low \emph{because of} \emph{mergers}.
To search for progenitors of our R-star mergers it would be best to
search for HeWD-BS binaries in high metallicity, old GCs, such as
those in the Galactic bulge e.g. NGC 6553 or 6528, \citep{2004MmSAI..75..398B,2004A&A...423..507Z},
with metallicities $[\mathrm{Fe}/\mathrm{H}]\sim-0.2$, or the LMC
clusters \citep{1986MNRAS.220..723E}.

The FK Com stars \citep{1981IAUS...93..177B,1981ApJ...247L.131B}
may be post-common envelope mergers: they are single, rapidly rotating
(close to or even beyond break-up e.g. \citealp{1993ApJ...404..316H,2006ApJ...644..464A}),
G/K-type giants which are thought to have evolved through a binary
merger \citep{1994ApJ...435..848W}. They are rare, about $2\times10^{-8}\,\mathrm{pc}^{-3}$
\citep{1982MNRAS.200..489C,1985AJ.....90..120H}, which is quite similar
to the R-star space density.

\subsection{Merger mixing and modelling }

\label{sub:Merger-mixing}In our merger scenario at $Z=0.02$ the
amount of carbon which is required to be mixed from the core into
the envelope during the helium flash is $0.003-0.01\mathrm{\, M_{\odot}}$
, assuming $\mathrm{C/O}=1$ in the R-star envelope. The analysis
of \citet{1984ApJS...55...27D} shows carbon and nitrogen are similarly
enhanced in {\rstar}s, with the $^{12}\mathrm{C}/^{13}\mathrm{C}$
ratio just above the CN-cycle equilibrium value of $4$. These observations
strongly suggest there is some hydrogen burning of the carbon-rich
material from the helium flash en route to the surface of the {\rstar}.
This would mean we have underestimated the amount of carbon dredge
up by a factor of about two because half the carbon is converted to
nitrogen. However, because some $^{13}\mathrm{C}$ and $^{14}\mathrm{N}$
may come from the deposition of material stripped from the primary
star during the first period of mass transfer, for some channels (e.g.
R3a) we require less CN burning. 

\refchange{Lithium poses a problem because the latest data of Zamora
(private communication, see also section \ref{sub:Spectroscopic-Studies})
suggest that {\rstar}s may be enhanced with lithium relative to normal
G/K giants. The fate of lithium during a stellar merger is not clear:
there will be some dilution and destruction, but perhaps also creation
by outward transport of $^{7}\mathrm{Be}$ in a way similar to that
of \citet{2000A&A...358L..49D}.}

\refchange{Hydrogen burning implies production of helium, which would
be mixed to the surface. However, hydrogen would have to be very deficient
to alter the spectrum and indeed this is not seen in the {\rstar}s.
In all our channels except R9 (the merger of two helium stars, which
has a negligible formation rate) a hydrogen envelope sits on top of
the merged core so the star would be observed as hydrogen rich.}

As discussed above, there may be a delay between the merger and helium
ignition. If the core can settle to hydrostatic equilibrium it should
be possible to model it with a one-dimensional stellar evolution code
which includes rotation and angular momentum transport (e.g. \citealp*{1998A&A...329..551L,2000ApJ...528..368H}
and \citealp[the series of papers by ][onward]{1997A&A...321..465M}).
The SPH models of \citet{2004A&A...413..257G} suggest that merged
cores settle down rapidly (minutes) compared to their nuclear burning
timescale (Myr) so the problem is tractable. Certainly, some one-dimensional
exploration of the problem would be useful.

This leads us to the realm of two and three dimensional modelling
of the core merger and associated mixing. Two dimensional simulations
without \emph{}rotation (\citealp{1980ApJ...239..284C,1981ApJ...247..607C,1987ApJ...317..724D,1996ApJ...471..377D})
do not show significant mixing of material from the core to the surface
-- yet we have good grounds, based on our population model and a lack
of alternatives, to suggest that it does happen. The only way forward
may be a full three-dimensional simulation of a merger of two HeWDs
inside a hydrogen-rich giant envelope \emph{including} nuclear burning.
This is some way off, and might not even be possible with current
methods such as SPH, but is perhaps not impossible in the near future
using codes such as Djehuty \citep{2006ApJ...639..405D}.

\subsection{Angular momentum}

\label{sub:Angular-momentum}In appendix \ref{sub:Merged-Core-Angular-Momentum}
we show that when two HeWD cores merge the velocity of the resultant
object exceeds its breakup velocity by a factor of about two, in agreement
with \citet{2006MNRAS.371.1381G}. There must be some form of angular
momentum transport out of the merged core, perhaps by magnetic fields
coupling the core to the envelope or shear mixing across the core-envelope
boundary (it has recently been suggested by \citealp{2007astro.ph..1528S}
that shear mixing may mix helium-rich, but not carbon-rich, material
out of the core of a red giant).

We show in appendix \ref{sub:Rotational-Velocity-of-postCE-object}
that if we conserve angular momentum and mass for the whole star,
after the merger it is rotating at about its breakup velocity. However,
it is not the case if some mass is lost during the common envelope
phase as this may remove enough angular momentum to stabilise the
star \citep*{1991ApJ...381..449D,Lovekin_Sills_2003}. It is still
rotating at a significant fraction of its breakup velocity, often
many tens of $\mathrm{km}\,\mathrm{s}^{-1}$. This contradicts the
observations of \citet{1997PASP..109..256M} which show that {\rstar}s
are \emph{not} rapidly rotating. There must be some angular momentum
loss, such as magnetic braking, which slows the {\rstar}, especially
because it contracts after helium ignites at the tip of the giant
branch.

\subsection{After the R-star\ldots{}}

\label{sub:After-the-R-star}The {\rstar}s are CHeB stars but eventually
they will exhaust their supply of helium and ascend the giant branch
again as $^{12}\mathrm{C}$-, $^{13}\mathrm{C}$- and $^{14}\mathrm{N}$-rich
AGB stars. They may be observed as J-type stars and indeed, perhaps
{\rstar}s are the progenitors of \refchange{some of} the J stars,
as suggested by \citet{1986MNRAS.220..723E}. If this is the case
then J stars should also be single stars, but it is not clear if this
is true. J stars are apparently 15\% of carbon-rich giants stars \citep{2000ApJ...536..438A}
and further work regarding their formation rate will determine if
{\rstar}s are the progenitors of some of them.

\section{Conclusions}

\label{sec:Conclusions}We have shown by means of a binary population
synthesis technique that binary mergers are a viable channel for the
formation of the {\rstar}s. Our models naturally reproduce \refchange{most of}
the properties of the {\rstar}s, namely that they are single, K-type
stars which show enhanced $^{12}\mathrm{C}$, $^{13}\mathrm{C}$ and
$^{14}\mathrm{N}$ without $s$-process or iron enhancement relative
to the sun.

We estimated the number of {\rstar}s from different binary merger
channels involving objects with helium cores. The most common merger
is that of a helium white dwarf with a red giant star and this makes
about ten times as many stars as we require if we are to match the
{\rstarabbrev} to red clump number ratio to the observed value. We
did, however, assume that the helium core of \emph{every} merger is
rotating and mixes carbon into the envelope upon helium ignition.
It is more likely that a small fraction of our stars ignite helium
while rotating rapidly enough to mix carbon into their envelope.

Other possible channels include mergers of a red giant with a Hertzsprung
gap star, a helium white dwarf with a Hertzsprung gap star or the
merger of two red giants. There must be other parameters which control
whether a merged star becomes a{\posta} {\rstar}, such as the core
mass or core-rotation rate. Our model {\rstar}s are rapidly rotating
core-helium burning giants so must undergo some kind of rapid magnetic
braking to slow them down, because observed {\rstar}s are not rotating
quickly.

As pointed out by \citet{1984ApJS...55...27D} and \citet{1997PASP..109..256M}
it is difficult to envisage an alternative evolutionary pathway which
leads to the {\rstar}s. Ten years on, we finally have a potential
explanation for their existence and their number. It is our hope that
this will stimulate further research in the area of helium-core mergers
inside common envelopes because for once we think we understand the
outcome of the merger process.

\begin{acknowledgements}
Our work on the {\rstar}s commenced at the {}``Nucleosynthesis in
Binary Stars'' workshop at the Lorentz centre, Leiden, 2005. RGI
is supported by the Nederlands Organisatie voor Wetenschappelijk Onderzoek
(NWO). He thanks Armagh Observatory and Monash University for funding
extended visits. He is grateful to the numerous people who have shared
their ideas with him over the past year on the subject of the paper,
especially Evert Glebbeek, Onno Pols and Pilar Gil-Pons for reading
the manuscript. CSJ acknowledges financial support from the Northern
Ireland Dept. of Culture Arts and Leisure (DCALNI), the UK Particle
Physics and Astronomy Research Council (PPARC) and the Netherlands
Research School for Astronomy (NOVA). Other aspects of this research
were supported by the DEST and the Australian Research Council. This
research has made use of the SIMBAD database, operated at CDS, Strasbourg,
France.
\end{acknowledgements}
\bibliographystyle{mn2e}

\appendix

\section{Appendix}

\label{sec:Appendix}

\subsection{Carbon Dredge Up}

\label{sub:Carbon-Dredge-Up}

We define abundances by mass fraction to be $X$ prior to the merger,
$Y$ post-merger. We then require

\begin{eqnarray}
\frac{4}{3}\frac{Y_{\mathrm{C}}}{Y_{\mathrm{O}}} & \geq & 1\end{eqnarray}
 to be a carbon star. The common envelope is of mass $M$ and the
amount of carbon required from the core is $\Delta_{\mathrm{C}}$
(a similar mass of material is mixed back down into the core). We
neglect the minor isotopes $^{13}\mathrm{C}$, $^{17}\mathrm{O}$
and $^{18}\mathrm{O}$ because their abundances are small (at most
$1/4$ of $\mathrm{C}$ is $^{13}\mathrm{C}$). The final mass of
carbon in the envelope is\begin{eqnarray}
Y_{\mathrm{C}}M_{\mathrm{env}} & = & X_{\mathrm{C}}(M-\Delta_{\mathrm{C}})+\Delta_{\mathrm{C}}\,,\end{eqnarray}
which gives

\begin{eqnarray}
Y_{\mathrm{C}} & = & \frac{X_{\mathrm{C}}(M-\Delta_{\mathrm{C}})+\Delta_{\mathrm{C}}}{M}\,.\end{eqnarray}
If we assume the oxygen abundance is unchanged, i.e. $Y_{\mathrm{O}}=X_{\mathrm{O}}$,
and there is no CN cycling of dredged up carbon, we have\begin{eqnarray}
Y_{\mathrm{C}} & \geq & \frac{3}{4}X_{\mathrm{O}}\,,\end{eqnarray}
which with the previous expression and some algebraic manipulation
gives us\begin{eqnarray}
\Delta_{\mathrm{C}} & \geq & \frac{\frac{3}{4}X_{\mathrm{O}}-X_{\mathrm{C}}}{1-X_{\mathrm{C}}}M_{\mathrm{}}\,.\end{eqnarray}
If a fraction $f$ of the dredged-up carbon is converted to nitrogen
as it passes through the hydrogen shell, the amount of dredged-up
carbon and nitrogen is \begin{eqnarray}
\Delta_{\mathrm{CN}} & \geq & \frac{1}{f}\frac{\frac{3}{4}X_{\mathrm{O}}-X_{\mathrm{C}}}{1-X_{\mathrm{C}}}M_{\mathrm{}}\,.\end{eqnarray}

\subsection{Merged Core Angular Momentum}

\label{sub:Merged-Core-Angular-Momentum}Here we consider the angular
momentum in the {\rstar} core as a result of the merger. When the
least massive helium core, of mass $m_{\mathrm{c}2}$, enters RLOF,
the orbital period is given by Kepler's law

\begin{eqnarray}
P & = & \frac{2\pi}{\sqrt{G}}\sqrt{\frac{a_{\mathrm{L}}^{3}}{m_{\mathrm{c1}}+m_{\mathrm{c}2}}}\,,\end{eqnarray}
where $a_{\mathrm{L}}$ is the separation and $m_{\mathrm{c}1}$ is
the mass of the more massive core. The equivalent orbital angular
velocity is then $\Omega=2\pi/P$. We can estimate the angular momentum
in the twin-core system $J$ by neglecting the spins of the stars
such that \begin{eqnarray}
J & = & J_{\mathrm{c}1}+J_{\mathrm{c}2}+J_{\mathrm{orb},\mathrm{c}}\nonumber \\
 & \simeq & J_{\mathrm{orb},\mathrm{c}}=\mu\Omega a_{\mathrm{L}}^{2}\end{eqnarray}
where $J_{\mathrm{c}1,2}$ are the spin angular momenta of the stars
and $\mu$ is the reduced mass. Then we have, by substitutions of
the expression for $\Omega$,\begin{eqnarray}
J_{\mathrm{orb,c}} & = & \frac{m_{\mathrm{c}1}m_{\mathrm{c}2}}{m_{\mathrm{c}1}+m_{\mathrm{c}2}}\Omega a_{\mathrm{L}}^{2}\nonumber \\
 & = & m_{\mathrm{c}1}m_{\mathrm{c}2}\sqrt{\frac{Ga_{\mathrm{L}}}{m_{\mathrm{c}1}+m_{\mathrm{c}2}}}\,.\label{eq:jorb_cores}\end{eqnarray}
Now, if we conservatively assume that no angular momentum is lost
from the cores during the merger then we can calculate the orbital
angular velocity $\omega$ of the merger product,\begin{eqnarray}
\omega & = & \frac{J_{\mathrm{orb,c}}}{km_{\mathrm{c}}r_{\mathrm{c}}^{2}}\nonumber \\
 & = & \frac{m_{\mathrm{c}1}m_{\mathrm{c}2}\sqrt{Ga\mathrm{_{\mathrm{L}}}}}{kr_{\mathrm{c}}^{2}\left(m_{\mathrm{c}1}+m_{\mathrm{c}2}\right)^{\frac{3}{2}}}\,,\end{eqnarray}
where $m_{\mathrm{c}}=m_{\mathrm{c}1}+m_{\mathrm{c}2}$ and $r_{\mathrm{c}}$
are the mass and radius of the merged core respectively and $k$ is
its radius of gyration. The velocity at the surface of the merged
core is \begin{eqnarray}
v_{\mathrm{F}} & = & \omega r_{\mathrm{c}}\nonumber \\
 & = & \frac{m_{\mathrm{c}1}m_{\mathrm{c}2}\sqrt{Ga\mathrm{_{\mathrm{L}}}}}{kr_{\mathrm{c}}\left(m_{\mathrm{c}1}+m_{\mathrm{c}2}\right)^{\frac{3}{2}}}\,.\end{eqnarray}
We can easily compare this to the break-up velocity 

\begin{eqnarray}
v_{\mathrm{B}} & = & \sqrt{\frac{Gm_{\mathrm{c}}}{r_{\mathrm{c}}}}\end{eqnarray}
to see \begin{eqnarray}
\frac{v_{\mathrm{F}}}{v_{\mathrm{B}}} & = & \frac{1}{k}\frac{m_{\mathrm{c}1}m_{\mathrm{c}2}}{\left(m_{\mathrm{c}1}+m_{\mathrm{c}2}\right)^{2}}\sqrt{\frac{a_{\mathrm{L}}}{r_{\mathrm{c}}}}\,.\end{eqnarray}
The separation at RLOF is $a_{\mathrm{L}}=\max\left[r_{\mathrm{c1}}/f(q),\, r_{\mathrm{c}2}/f(1/q)\right]$
where $f$ is given by the function of \citet{1983ApJ...268..368E},
and given $m_{\mathrm{c}1}$ and $m_{\mathrm{c}2}$ we obtain the
core radii from the formulae of \citet{1997MNRAS.291..732T}. We set
$k=0.21$ as for an $n=3/2$ polytrope. For $0.1\leq m_{\mathrm{c}1}=m_{\mathrm{c}2}\leq0.5$
the ratio $1.9\lesssim v_{\mathrm{F}}/v_{\mathrm{B}}\lesssim2.7$
so there must be some form of outward angular momentum transport in
order for the cores to merge.

\subsection{Rotational Velocity of the Post-CE Star}

\label{sub:Rotational-Velocity-of-postCE-object}We can derive the
maximum velocity of the star which results from common envelope evolution
by conserving angular momentum during the merger. At the beginning
of the CE phase the binary system has a total angular momentum $J=J_{\mathrm{orb}}+J_{1}+J_{2}$
where $J_{1}$ and $J_{2}$ are the spin angular momenta of the stars
and $J_{\mathrm{orb}}$ is the orbital angular momentum. The \emph{}final
angular velocity $\omega$ is calculated from a solid-body approximation,
assuming angular momentum is conserved during the merger, \begin{eqnarray}
J & = & (k_{2}M_{\mathrm{env}}+k_{3}M_{\mathrm{c}})R^{2}\omega\end{eqnarray}
where $M_{\mathrm{env}}$ , $M_{\mathrm{c}}$, $R$ are the envelope
and core mass and the radius of the merged star respectively. The
constant $k_{3}=0.21$ as previously, but $k_{2}$ depends on the
structure of the giant star (it is about $0.15-0.2$). We then solve
for the velocity $v=\omega R$ and compare to the breakup velocity
$v_{\mathrm{B}}$ (c.f. section \ref{sub:Merged-Core-Angular-Momentum}).
For a typical R3a system, with initial parameters $M_{1}=1.25\mathrm{\, M_{\odot}}$,
$M_{2}=0.5\mathrm{\, M_{\odot}}$, $P=1.6\,\mathrm{days}$, $Z=0.02$,
$\alpha_{\mathrm{CE}}=1$, $\lambda_{\mathrm{CE}}=0.5$, $e=0$, no
extra CRAP, we find the merged stellar mass is $1.34\mathrm{\, M_{\odot}}$
($0.05\mathrm{\, M_{\odot}}$ is lost during the common envelope phase,
we have \emph{not} taken into account the loss of angular momentum
due to this mass) with a core of mass $0.41\mathrm{\, M_{\odot}}$
and $k_{2}=0.143$, hence $v=125\,\mathrm{km}/\mathrm{s}$, about
$95\%$ of the break up velocity ($132\,\mathrm{km}/\mathrm{s}$).
The ratio reduces to $60\%$ if we assume the angular momentum in
the cores does not contribute to the rotation of the envelope, which
is required for the core to be rapidly spinning when it ignites helium.
If we attempt to take into account the angular momentum that is lost
in the $0.05\mathrm{\, M_{\odot}}$ of material ejected during the
common envelope phase, presumably at approximately the breakup velocity
so $\Delta J=v_{\mathrm{B}}R\times0.05\mathrm{\, M_{\odot}}$, the
ratio $v/v_{\mathrm{B}}$ drops further to $28\%$ ($v=37\,\mathrm{km}/\mathrm{s}$).
It would seem these systems are, in general, quite rapidly rotating
(at least tens of $\mathrm{km}/\mathrm{s}$), but are not all at the
breakup velocity. 
\end{document}